# Variational Calculation of Relativistic Meson-Nucleon Scattering In Zeroth Order

A. W. Schreiber [1,2], R. Rosenfelder [1] and C. Alexandrou [1,3]

[1] Paul Scherrer Institute, CH-5232 Villigen PSI, Switzerland
[2] TRIUMF, 4004 Wesbrook Mall, Vancouver, B.C., Canada V6T 2A3
[3] Department of Natural Sciences, University of Cyprus, CY-1678 Nicosia, Cyprus

**Abstract**

We extend the polaron variational treatment previously developed for the propagator to the case where one nucleon and $n$ external mesons are present. Using the particle representation of the scalar Wick-Cutkosky model this is done in lowest order of an expansion of the exact action around a retarded quadratic trial action. In particular, we evaluate the form factor for scattering of mesons from the scalar nucleon and determine the radius of the dressed particle. After analytic continuation to Minkowski space we study elastic meson-nucleon scattering both analytically and numerically near threshold and show that it is essential to incorporate the correct behaviour of the retardation function at large proper times. Only if this is done the optical theorem is approximately fulfilled over a range of energies and coupling constants.

# 1 Introduction

This is the third paper in a series which explores the use of variational principles in the particle representation of field theory. Following Feynman's treatment of the polaron [1] we have applied similar techniques [2, 3] to the simplest scalar field theory which involves a Yukawa interaction of heavy particles ('nucleons') with light mesons. This is the Wick-Cutkosky model [4, 5] described by the following Lagrangian in euclidean space time

$$\mathcal{L} = \frac{1}{2} \left( \partial_\mu \Phi \right)^2 + \frac{1}{2} M_0^2 \Phi^2 + \frac{1}{2} \left( \partial_\mu \varphi \right)^2 + \frac{1}{2} m^2 \varphi^2 - g \Phi^2 \varphi \tag{1}$$

Here $M_0$ is the bare mass of the heavy particle $\Phi$, $m$ is the mass of the light particle $\varphi$ and $g$ is the (dimensionfull) coupling constant of the Yukawa interaction between the two particles. In the present work we will concentrate on the Green function for *one* nucleon interacting with an *arbitrary* number of mesons. After integrating out the mesons (which is possible if the nucleons are quenched) we obtained in (I, II) the following generating functional for the corresponding (connected) Green functions

$$Z[j, x] = \text{const.} \int_0^\infty d\beta \, \exp\left(-\frac{\beta}{2} M_0^2\right) \int_{x(0)=0}^{x(\beta)=x} \mathcal{D}x(\tau) \, \exp\left(-S_{\text{eff}}[x(\tau)] - S_2[x(\tau), j]\right). \tag{2}$$

Here the effective action for the nucleon is given by

$$S_{\text{eff}}[x(\tau)] = \int_0^\beta d\tau \, \frac{1}{2} \dot{x}^2 - \frac{g^2}{2} \int_0^\beta d\tau_1 \int_0^\beta d\tau_2 \int \frac{d^4q}{(2\pi)^4} \frac{1}{q^2 + m^2} e^{iq \cdot (x(\tau_1) - x(\tau_2))} \tag{3}$$

and the meson source terms are contained in

$$S_2[x(\tau), j] = -g \int d^4y \, j(y) \int_0^\beta d\tau \int \frac{d^4q}{(2\pi)^4} \frac{e^{iq \cdot (y - x(\tau))}}{q^2 + m^2}. \tag{4}$$

Eqs. (2 - 4) define the "particle representation" of the Wick-Cutkosky model in the 'proper-time gauge' [7] for the sector of the theory which we consider here: they are formulated in terms of trajectories $x(\tau)$ of the nucleon which are parametrized by the proper time $\tau$ and obey the boundary conditions $x(0) = 0$ and $x(\beta) = x$. To obtain the Green functions for one nucleon (propagating from 0 to $x$) and $n$ external mesons one has to do the usual differentiation with respect to the meson sources $j(y)$, perform the remaining path integral over all trajectories of the nucleon and finally integrate over $\beta$ from zero to infinity with the weight $\exp(-\beta M_0^2/2)$. It is, of course, impossible to perform this path integral exactly. In (I, II) we have therefore approximated it variationally on the pole of the 2-point function by a retarded quadratic two-time action. Various parameterizations for the retardation function which enters this trial action have been investigated and for the most general case we have solved the variational equations numerically. This was possible for values of the dimensionless coupling constant

$$\alpha = \frac{g^2}{4\pi M^2} \leq 0.815, \tag{5}$$

where $M$ is the physical mass of the nucleon. For $\alpha$ greater than the critical value $\alpha_c = 0.815$ no real solutions could be obtained. The emergence of a critical coupling constant in our variational calculation points to the well-known instability present in the Wick-Cutkosky model [6].

It is the purpose of the present paper to extend the treatment given in (I, II) for the 2-point function to processes which involve $n$ external mesons. We will do this in 'zeroth' order in the difference between the exact effective action (3) and the trial action

$$\Delta S = S_{\text{eff}} - S_t, \tag{6}$$

i.e. we will utilize the trial action $S_t$ which has been determined previously for the 2-point function to evaluate the $(2+n)$-point function. We will demonstrate that already this lowest order gives reasonable and non-trivial results. In particular, we will study in detail meson-nucleon scattering in this model which requires an analytic continuation of our results to Minkowski space. Remarkably this was already anticipated by K. Mano [8] who first pioneered the use of polaron variational methods in the Wick-Cutkosky model.

This paper is organized as follows: in Section 2 we recall some elements of the polaron variational approach which we will need in the present paper. In Section 3 we consider the case where there are $n$ external mesons and derive the lowest order approximation to the corresponding $(2+n)$-functions in an expansion of the exact effective action around the trial action. Section 4 discusses the special case of $G_{2,2}$, the 'Compton' amplitude, the analytic continuation and the correct form of the trial action. In Section 5 we present our numerical results for differential and total cross sections near threshold. The main results of this work are summarized in the last Section, whereas some technical details are relegated to the Appendix.

## 2 Polaron Variational Approach

Following Feynman's treatment of the polaron problem we have performed in (I) a variational calculation of the 2-point function with the quadratic trial action

$$S_t[x] = \int_0^\beta d\tau \, \frac{1}{2} \dot{x}^2 + \int_0^\beta d\tau_1 \int_0^{\tau_1} d\tau_2 \, f(\tau_1 - \tau_2) \, [\, x(\tau_1) - x(\tau_2) \,]^2 \, . \tag{7}$$

Here $f(\tau_1 - \tau_2)$ is an undetermined 'retardation function' which takes into account the (proper) time lapse occurring when mesons are emitted and absorbed on the nucleon. In actual calculations it is more convenient to use the Fourier space form

$$S_t[b] = \sum_{k=0}^\infty A_k \, b_k^2 \, , \tag{8}$$

where the $b_k$ are the Fourier components of the path $x(\tau)$ and the Fourier coefficients $A_k$ are either parametrized or considered as variational parameters. Eq. (8) contains as a special case the free action for which all $A_k$ equal unity. Near the nucleon pole the 2-point function should behave like [1]

$$G_{2,0}(p^2) \longrightarrow \frac{Z}{p^2 + M^2} \tag{9}$$

where $0 < Z < 1$ is the residue. As was shown in (I) this requires the proper time $\beta$ to tend to infinity. For $\beta \to \infty$ all discrete sums over Fourier modes $A_k$ turn into integrals over the 'profile function' $A(E = k\pi/\beta)$. Mathematically this is achieved by using Poisson's summation formula [9] which, for an even function $F(k\pi/\beta)$, reads

$$\sum_{k=1}^\infty F\left(\frac{k\pi}{\beta}\right) = \frac{\beta}{\pi} \int_0^\infty dE \, F(E) - \frac{1}{2} F(0) + \frac{2\beta}{\pi} \sum_{n=1}^\infty \int_0^\infty dE \, F(E) \, \cos(2n\beta E) \, . \tag{10}$$

For $\beta \to \infty$ the terms in the last sum are exponentially suppressed.

The profile function $A(E)$ is linked to the retardation function $f(\sigma)$ through

$$A(E) = 1 + \frac{8}{E^2} \int_0^\infty d\sigma \, f(\sigma) \, \sin^2 \frac{E\sigma}{2} \, . \tag{11}$$

---
[1] In this work all 4-momenta are taken to be euclidean.

In Refs. (I, II) we have studied the *'Feynman' parametrization* given by

$$f_F(\sigma) = w \frac{v^2 - w^2}{4} e^{-w\sigma} , \qquad (12)$$

which leads to

$$A_F(E) = \frac{v^2 + E^2}{w^2 + E^2} , \qquad (13)$$

and an *'improved' parametrization*

$$f_I(\sigma) = \frac{v^2 - w^2}{2w} \frac{1}{\sigma^2} e^{-w\sigma} , \qquad (14)$$

which gives

$$A_I(E) = 1 + 2 \frac{v^2 - w^2}{wE} \left[ \arctan \frac{E}{w} - \frac{w}{2E} \ln \left( 1 + \frac{E^2}{w^2} \right) \right] . \qquad (15)$$

In both cases $v, w$ are parameters whose values have to be determined by minimizing the variational functional given in Eq. (27) of (II). As the bare mass $M_0$ requires renormalization and the physical mass $M = 939$ MeV is fixed by the pole position of the propagator, the minimum does not have a physical meaning but only tells us how good the variational ansatz is. The numerical calculations in (II) have shown that the 'improved' parametrization is significantly better than Feynman's because it includes the correct short-time behaviour. Note also that the singularity structure of the corresponding profile functions in the complex $E$-plane is quite different: whereas $A_F(E)$ exhibits poles at $E = \pm iw$, the profile function of the 'improved' parametrization has *branch points* at $E = \pm iw$.

Within the gaussian ansatz the best approximation is obtained by not imposing any specific form for the retardation function but determining it from varying the variational functional with respect to the profile function $A(E)$. As shown in (I) this gives the following expression for the 'variational' retardation function

$$f_{\text{var}}(\sigma) = \frac{g^2}{32\pi^2} \frac{1}{\mu^4(\sigma)} \int_0^1 du \left[ 1 + \frac{m^2}{2} \mu^2(\sigma) \frac{1-u}{u} - \frac{\lambda^2 M^2 \sigma^2}{2\mu^2(\sigma)} u \right] e \left( m\mu(\sigma), \frac{\lambda M \sigma}{\mu(\sigma)}, u \right) . \qquad (16)$$

Here $0 < \lambda < 1$ is a variational parameter determined by solving the coupled system of variational equations. In addition, we have used the abbreviations

$$e(a, b, u) = \exp \left( -\frac{a^2}{2} \frac{1-u}{u} - \frac{b^2}{2} u \right) \qquad (17)$$

and

$$\mu^2(\sigma) = \frac{4}{\pi} \int_0^\infty dE \frac{1}{A(E)} \frac{\sin^2(E\sigma/2)}{E^2} . \qquad (18)$$

We call $\mu^2(\sigma)$ the 'pseudotime' because

$$\mu^2(\sigma) \xrightarrow{\sigma \to 0} \sigma \qquad (19)$$

and

$$\mu^2(\sigma) \xrightarrow{\sigma \to \infty} \frac{\sigma}{A_0} \qquad (20)$$

where $A_0 \equiv A(0)$ is the value of the profile function at $E = 0$. It should be noted that this variational retardation function has the same $1/\sigma^2$-behaviour for small relative times as the 'improved' parametrization (14).

# 3  Green Functions with n External Mesons

In the previous work we have been considering the propagator of the nucleon dressed by (virtual) mesons and have used this Green function to fix the variational parameters appearing in the trial action. It is of course possible to apply the same procedure for any other Green function as well. Alternatively, one can estimate other Green functions by evaluating them with the variational parameters determined from the nucleon propagator. We shall follow this procedure for the general Green function with two external nucleon and $n$ meson legs which we shall call the $(2+n)$-point function. With the help of the reduction formulas these Green functions describe the matrix elements for processes like meson absorption on a nucleon, meson-nucleon scattering or nucleon-antinucleon annihilation in our simplified scalar model. Nonperturbative production of multiboson states has attracted a lot of theoretical interest recently (see e.g. [10]), in a different context.

For simplicity we will work to zeroth order in $\langle \Delta S \rangle$ and within the framework of the 'coordinate averaging' described in (I). The general expression for the $(2+n)$-point function in coordinate space may be obtained from the generating functional for connected Green functions (Eq. (2)) by repeated differentiation with respect to the source $j(y)$. After Fourier transformation we may write

$$G_{2,n}(p,p';\{q\}) = \text{const.} \int_0^\infty d\beta \exp\left[-\frac{\beta}{2}M_0^2\right] \int d^4x\, e^{-ip'\cdot x}$$
$$\cdot \int_{x(0)=0}^{x(\beta)=x} \mathcal{D}x(\tau) \prod_{i=1}^n \left[g \int_0^\beta d\tau_i e^{iq_i\cdot x(\tau_i)}\right] e^{-S_{\text{eff}}[x(\tau)]}, \qquad (21)$$

where the external meson propagators as well as an overall momentum conserving delta function have been removed, while the truncation of the external nucleon propagators still remains to be performed. Our convention is that nucleon lines are ingoing with momenta $p$ and $p'$ whereas mesons are outgoing with momenta $q_i$. The effective action $S_{\text{eff}}$ for the nucleon is given in Eq. (3). We shall determine the normalization of Eq. (21) later. The proper time $\tau_i$ appearing in Eq. (21) may be interpreted as the time at which the meson with momentum $q_i$ couples to the nucleon, the latter being at the space-time point $x(\tau_i)$.

To zeroth order in $\langle \Delta S \rangle$, $G_{2,n}(p,p';\{q\})$ is given by Eq. (21), with $S_{\text{eff}}$ replaced by $S_t$. Using the definition for $S_t$ in Eq. (8) one may do the integral over the endpoint

$$\int d^4x \exp\left[-ip'\cdot x - A_0\frac{x^2}{2\beta} + i\sum_{i=1}^n \frac{\tau_i}{\beta}q_i\cdot x\right] = \left[\frac{2\pi\beta}{A_0}\right]^2 \exp\left[-\frac{1}{2A_0\beta}\left(\sum_{i=1}^n q_i\tau_i - p'\beta\right)^2\right], \qquad (22)$$

as well as the path integral

$$\int_{x(0)=0}^{x(\beta)=x} \mathcal{D}x(\tau) \exp\left[\sum_{k=1}^\infty \left(i\lambda_k'\cdot b_k - A_k b_k^2\right)\right] = \frac{\text{const.}}{(2\pi\beta)^2} \exp\left[\sum_{k=1}^\infty \frac{\lambda_k'^2}{4A_k}\right] \prod_{k=1}^\infty \frac{1}{A_k^2} \qquad (23)$$

where

$$\lambda_k' = \frac{2\sqrt{\beta}}{\pi k} \sum_{i=1}^n \sin\frac{k\pi\tau_i}{\beta} q_i \ . \qquad (24)$$

One may perform the sum appearing on the right hand side of Eq. (23) using the summation formula (10). One obtains, up to exponentially small terms in $\beta$,

$$\sum_{k=1}^\infty \frac{\lambda_k'^2}{4A_k} = -\frac{1}{2\beta A(0)}\left(\sum_{i=1}^n \tau_i q_i\right)^2 + \frac{1}{2A(0)}\sum_{i,j=1}^n q_i\cdot q_j \min(\tau_i,\tau_j)$$
$$+ \frac{1}{\pi}\sum_{i,j=1}^n q_i\cdot q_j \int_0^\infty \frac{dE}{E^2}\left(\frac{1}{A(E)} - \frac{1}{A(0)}\right)\sin(E\tau_i)\sin(E\tau_j) \ . \qquad (25)$$

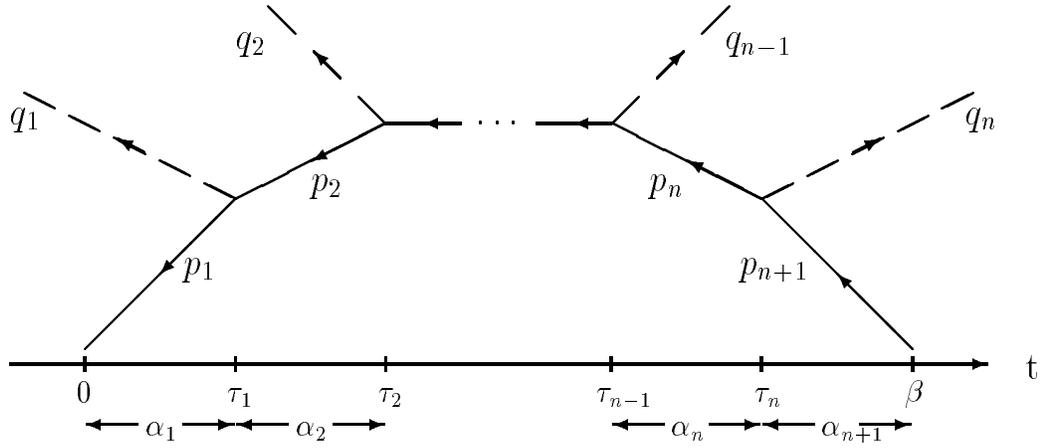

Figure 1: The definition of the relative times and momenta. Note that $\tau_0 = 0$, $\tau_{n+1} = \beta$, $p_1 = -p$ and $p_{n+1} = p'$.

Note that the first term, which is quadratic in the proper times $\tau$ and therefore potentially troublesome, will cancel against the identical term appearing in Eq. (22).

The expression (21) for the $n$-point function involves an integral over all possible orderings of the times $\{\tau_i\}$. Equivalently, we may relabel the integration variables such that they correspond to the particular ordering $0 \leq \tau_1 \leq \tau_2 \ldots \leq \tau_n \leq \beta$ and sum over all possible permutations of the momenta $\{q_i\}$. Furthermore, it is more sensible to use relative rather than absolute times, so we shall define (see Fig. 1)

$$\begin{aligned} \alpha_1 &= \tau_1 \\ \alpha_2 &= \tau_2 - \tau_1 \\ &\vdots \\ \alpha_n &= \tau_n - \tau_{n-1} \\ \alpha_{n+1} &= \beta - \tau_n \;, \end{aligned} \tag{26}$$

so that

$$\tau_i = \sum_{k=1}^{i} \alpha_k \qquad \text{and} \qquad \beta = \sum_{k=1}^{n+1} \alpha_k \;. \tag{27}$$

The Jacobian of this transformation is one, and the product of integrals simplifies to

$$\int_0^\infty d\beta \int_0^\beta d\tau_n \int_0^{\tau_n} d\tau_{n-1} \ldots \int_0^{\tau_2} d\tau_1 = \prod_{i=1}^{n+1} \int_0^\infty d\alpha_i \;. \tag{28}$$

With this transformation the terms linear in the times appearing in the exponential become

$$-\frac{1}{2A_0}\left[\beta\, p'^2 - 2p' \cdot \sum_{i=1}^{n} q_i \tau_i + \sum_{i,j=1}^{n} q_i \cdot q_j\, \min(\tau_i, \tau_j)\right] = -\frac{1}{2A_0} \sum_{i=1}^{n+1} \alpha_i\, p_i^2 \;, \tag{29}$$

where the momenta $p_i$ are defined in Fig. 1 and we have used the fact that the overall momentum is conserved. We are thus led to the following expression for the $(2+n)$-point function:

$$G_{2,n}(p,p';\{q\}) = \frac{1}{2\,g} \sum_{\mathcal{P}(\{q\})} \prod_{i=1}^{n+1} \left\{ g \int_0^\infty d\alpha_i\, \exp\left[-\frac{\alpha_i}{2A_0}(p_i^2 + A_0 M_0^{\,2})\right] \right\} \prod_{k=0}^{\infty} \left\{\frac{1}{A_k^{\,2}}\right\}$$

$$\cdot \exp\left\{-\frac{1}{\pi}\sum_{i,j=1}^{n} q_i \cdot q_j \int_0^\infty \frac{dE}{E^2}\left[\frac{1}{A(E)} - \frac{1}{A(0)}\right]\right. \tag{30}$$

$$\left. \cdot \sin\left(E\sum_{l=1}^{i}\alpha_l\right)\sin\left(E\sum_{l=1}^{j}\alpha_l\right)\right\} \quad .$$

This expression is valid for all non-negative integers $n$ and, as we shall see, is properly normalized. Before treating the case where $n \ne 0$, let us first consider what the result is for the nucleon propagator at this order.

### 3.1 $n = 0$

Applying the Poisson summation formula (10) to the infinite product of Fourier coefficients gives

$$\prod_{k=0}^{\infty} \frac{1}{A_k^2} = \frac{1}{A_0}\exp\left[-\frac{2\beta}{\pi}\int_0^\infty dE\, \log A(E) + \text{Ex}(\beta)\right] \quad . \tag{31}$$

The remainder

$$\text{Ex}(\beta) = 2\sum_{n=1}^{\infty}\frac{1}{n\pi}\int_0^\infty dE\, \frac{A'(E)}{A(E)}\sin(2n\beta E) \tag{32}$$

is exponentially suppressed for large $\beta$. Hence close to the nucleon pole, i.e. for $\beta \to \infty$, we obtain

$$G_{2,0}(p,p) = \left[p^2 + A_0 M_0^2 + \frac{4A_0}{\pi}\int_0^\infty dE\, \log A(E)\right]^{-1} \quad . \tag{33}$$

So at zeroth order the residue of the nucleon propagator is one

$$Z^{(0)} = 1 \tag{34}$$

and the physical mass is given by

$$M^2 = A_0\left[M_0^2 + \frac{4}{\pi}\int_0^\infty dE\, \log A(E)\right] \quad . \tag{35}$$

Note that the singular behaviour of the 'improved' and the 'variational' retardation functions for small $\sigma$ leads to an $1/E$ fall-off of the corresponding profile function for large $E$. This in turn requires (even in zeroth order) an infinite renormalization of the bare mass $M_0$ since the $E$-integral in Eq. (35) does not converge.

As we have already noted, Eq. (33) is only valid near the pole. It is usually not possible to give a closed expression for the infinite product in Eq. (31) or the infinite sum in Eq. (32) which is also valid for subasymptotic $\beta$'s. One case, however, where this may be carried out exactly is for the Feynman parameterization given in Eq. (13), in which case we have

$$\prod_{k=0}^{\infty}\frac{1}{A_k^2} = \left[\frac{w\,\sinh w\beta}{v\,\sinh v\beta}\right]^2 \tag{36}$$

so that the propagator becomes

$$G_{2,0}(p,p) = \frac{w^2}{2\,v^2}\int_0^\infty d\beta\,\left[\frac{\sinh w\beta}{\sinh v\beta}\right]^2 \exp\left[-\frac{\beta}{2A_0}(p_i^2 + A_0 M_0^2)\right] \quad . \tag{37}$$

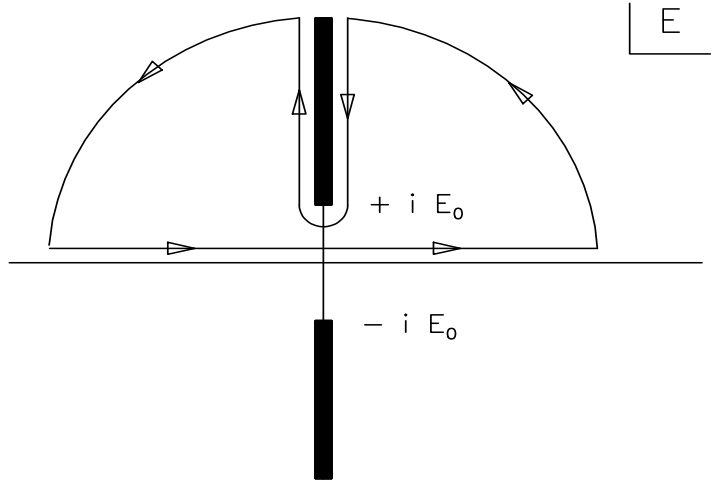

Figure 2: Singularity structure and integration contours for the 'improved' profile function in the complex $E$-plane.

One may even carry out this last integral over $\beta$ to obtain

$$G_{2,0}(p,p) = \frac{w^2}{4\,v^3}\left[\left(a+\frac{2w}{v}-1\right)\Psi\left(a+\frac{2w}{v}\right)+(a-1)\,\Psi(a)\right.$$
$$\left.-2\left(a+\frac{w}{v}-1\right)\Psi\left(a+\frac{w}{v}\right)\right] \tag{38}$$

with $a = w^2(p^2+M^2)/(4v^3)$. Here $\Psi(x)$ is the digamma function [11], which has simple poles at $x = 0, -1, -2, \ldots$.

It is clear that this Green function has the unfortunate feature of having not only the pole situated at the nucleon mass, but also an infinite sequence of 'ghost poles' (unless $v = w$, which corresponds to $g = 0$, in which case all ghost poles cancel). Note that the residues at these ghost poles cancel each other, which is consistent with $Z^{(0)} = 1$ noted above. The closest pole is situated at $p^2+M^2 = -4v^2/w$, which numerically turns out to be more than about $(2M)^2$ away from the nucleon pole. Hence, for the purpose of describing low-energy processes, these ghosts seem not to be relevant physically.

What happens to these ghost poles for a general profile function $A(E)$? Without a specific parametrization it is possible to investigate the nearest singularities of the zeroth order propagator by keeping only the $n = 1$ term in the sum for the remainder (32) and expanding the exponential. This gives

$$G_{2,0}(p,p) \simeq \frac{1}{p^2+M^2} + \frac{8A_0}{\pi}\int_0^\infty dE\, E\, \frac{A'(E)}{A(E)}\frac{1}{(p^2+M^2)^2+(4A_0 E)^2}\,. \tag{39}$$

To make further progress we need some knowledge of the analytic behaviour of $A(E)$ in the complex $E$-plane. It has already been observed that the 'improved' parametrization of the profile function has branch cuts on the imaginary axis starting at $\pm iw$. Assuming this structure to hold in general (with the cuts starting at some value $\pm iE_0$) we can, after making use of the fact that the profile function is even in $E$, deform the integration contour as shown in Fig. 2 and obtain

$$G_{2,0}(p,p) \simeq \frac{1}{p^2+M^2} + \int_{E_0}^\infty dE\, \frac{r(E)}{p^2+M^2+4A_0 E} \tag{40}$$

with

$$r(E) = \frac{2}{\pi} \text{Im} \frac{A'(iE+\epsilon)}{A(iE+\epsilon)} \, . \tag{41}$$

Eq. (40) has the form of a Källén-Lehmann representation [12] for the euclidean propagator and shows that the ghost poles in Feynman parametrization turn into branch points of the off-shell propagator - at least for the 'improved' parametrization. It is tempting to identify these branch points with the beginning of the meson production cuts which should be present in the exact 2-point function. However, the threshold at $p^2 + M^2 = -4A_0 E_0$ turns out to be too high if the numerical values of $A_0$ and $E_0 = w$ tabulated in (II) are inserted, and, more seriously, the weight function $r(E)$ is not positive definite. The latter deficiency is also obvious from the result

$$\int_{E_0}^\infty dE \, r(E) = \frac{1}{i\pi} \int_{E_0}^\infty dE \, \frac{d}{dE} \left[ \ln A(iE + \epsilon) - \ln A(iE - \epsilon) \right] = 0 \, . \tag{42}$$

Although this is fully consistent with the general sum rule for the weight function in the Källén-Lehmann representation [12] and $Z^{(0)} = 1$, it means that, in general, we have 'ghost branch points' instead of ghost poles. These unwanted properties are not as disastreous as it seems at first sight: it should be remembered that we have *extrapolated* away from the nucleon pole with the variational parameters fixed to their on-shell value. As shown in (II) the off-shell variational equations necessarily lead to a dependence of the variational parameters on the virtuality $p^2 + M^2$ which will cure these deficiencies. We will not pursue this in the present work but will keep the variational parameters as determined on the pole of the 2-point function.

### 3.2 $n \neq 0$

Having determined the residue of the nucleon propagator at zeroth order, we may now check that the Green function (30) is properly normalized by considering the small coupling limit. In order to obtain the lowest order contribution in the coupling $g$, we just need to set the profile function $A(E)$ equal to one and $M_0$ to $M$. One can now do the integrals over the times $\alpha_i$ to obtain

$$G_{2,n}^{\text{tree}}(p, p'; \{q\}) = \frac{1}{2g} \sum_{\mathcal{P}(\{q\})} \prod_{i=1}^{n+1} \frac{2g}{p_i^2 + M^2} \tag{43}$$

as required.

Let us now specialize to the case where the nucleon is on-shell. As discussed in (I) this corresponds to considering the integration region where the corresponding time differences $\alpha_1$ and $\alpha_{n+1}$ go to infinity. Hence, using again the Poisson summation formula, the terms quadratic in the meson momenta may be simplified by writing the product of the two sine functions as a sum over two cosines, one of which only gives rise to exponentially small (in the variable $\alpha_1$) terms. One obtains

$$\frac{1}{\pi} \int_0^\infty \frac{dE}{E^2} \left[ \frac{1}{A(E)} - \frac{1}{A(0)} \right] \sin\left( E \sum_{l=1}^i \alpha_l \right) \sin\left( E \sum_{l=1}^j \alpha_l \right) \stackrel{\alpha_1 \to \infty}{\Longrightarrow} \xi\left( \sum_{l=\min(i,j)+1}^{\max(i,j)} \alpha_l \right) \, , \tag{44}$$

where

$$\xi(\tau) = \frac{1}{2\pi} \int_0^\infty \frac{dE}{E^2} \left[ \frac{1}{A(E)} - \frac{1}{A(0)} \right] \cos(E\tau) \, . \tag{45}$$

Note that the argument of $\xi$ in Eq. (44) is zero if $i = j$. Furthermore, $\tau$ does not depend on $\alpha_1$ or $\alpha_{n+1}$ for any values of $i$ and $j$ and so we may now carry out the integrations over these two times, yielding the propagators for the external nucleon legs. The $(2+n)$-point Green function for on-shell nucleons may therefore be written as

$$G_{2,n}(p,p';\{q\}) = \frac{1}{p^2+M^2}\frac{1}{p'^2+M^2} G_{2,n}^{\text{tr}}(p,p';\{q\}) \quad , \tag{46}$$

where

$$\begin{aligned} G_{2,n}^{\text{tr}}(p,p';\{q\}) &= 2g\, A_0 \sum_{\mathcal{P}(\{q\})} \prod_{i=2}^{n} \left\{ g \int_0^\infty d\alpha_i \, \exp\left[-\frac{\alpha_i}{2A_0}(p_i{}^2+M^2)\right] \right\} \\ &\cdot \exp\left\{-\sum_{i,j=1}^{n} q_i \cdot q_j\, \xi\left(\sum_{l=\min(i,j)+1}^{\max(i,j)} \alpha_l\right)\right\} \end{aligned} \tag{47}$$

and we have used the zeroth order expression Eq. (35) for the physical mass.

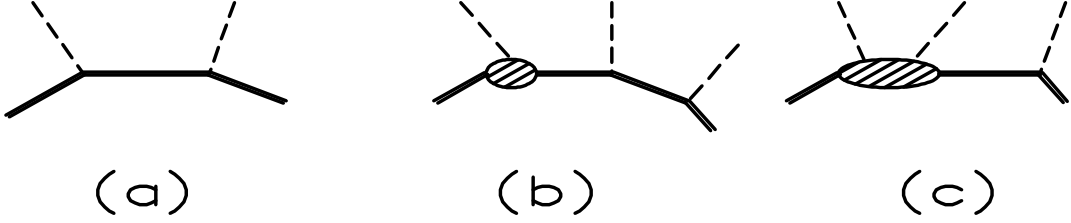

Figure 3: Some diagrams included in the $(2+n)$-point function. Each solid line stands for a fully dressed nucleon propagator.

Note that all momenta in Eq. (47) appear quadratically in the exponent. This is due to the fact that it corresponds to the zeroth order term in a variational calculation with a quadratic trial action. Nevertheless, already at this order Eq. (47) contains a remarkable amount of information. To see this, it is instructive to examine its behaviour in perturbation theory. It suffices to expand the exponential in the $q_i$'s. The leading term corresponds precisely to the tree diagram already encountered before (Eq. (43)), with fully dressed nucleon propagators for the internal lines and bare vertices for the external pions. An example of this type of term is shown in Fig. 3 a. Next, there is a class of diagrams containing only diagonal terms ($i=j$). These correspond to similar diagrams, however this time with one or more dressed vertices (e.g. Fig. 3 b). This class of diagrams is reducible on each side of each dressed vertex. This is due to the fact that $\xi(0)$ is independent of the $\alpha$'s and so the integrals over the times may be carried out, yielding a product of propagators and form factors. Furthermore, there is a class of diagrams where off-diagonal terms ($i\neq j$) are also present. Because for these terms $\xi$ contains at least one $\alpha$, these correspond to diagrams which are not reducible on each internal leg (e.g. Fig. 3 c). For example the term in the expansion with $i=1$ and $j=2$ gives $q_1 \cdot q_2\, \xi(\alpha_2)$ and thus corresponds to a diagram with meson exchange across the first and second vertex. Lastly, note that the exponential gives rise to any combination of the above, to all orders of the coupling.

The function $\xi(\tau)$ which we have introduced in Eq. (45) is closely related to the pseudotime $\mu^2(\tau)$. Indeed using Eq. (18) we obtain

$$\xi(\tau) = \xi(0) + \frac{\tau}{4A(0)} - \frac{1}{4}\mu^2(\tau)\,. \tag{48}$$

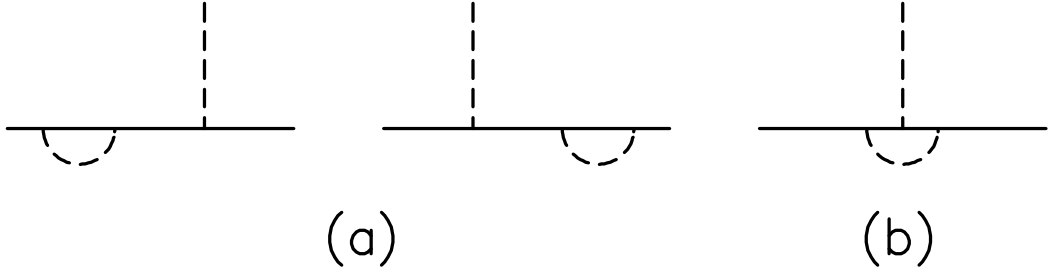

Figure 4: Third-order graphs for $G_{2,1}$ : (a) self-energies, (b) vertex correction.

Note, in particular, that
$$\frac{1}{A_0} = 1 + 4\,\xi'(0) \tag{49}$$
so that all on-shell Green functions are completely determined by the function $\xi(\tau)$. For the Feynman parametrization (13) we have the analytical result
$$\xi_F(\tau) = \frac{v^2 - w^2}{4v^3}\,e^{-v\tau}\;, \tag{50}$$
which illustrates the fact that $\xi(\tau)$ is an exponentially decaying function of $\tau$ (see Eq. (A.27)).

Finally, for convenience we explicitly write down the lowest order Green functions, where all of the above points may be illustrated :
$$G_{2,1}^{\mathrm{tr}}(p,p';q) = 2g\,A_0\,e^{-q^2\,\xi(0)} \tag{51}$$
$$G_{2,2}^{\mathrm{tr}}(p,p';q_1,q_2) = 2g^2\,A_0\,e^{-(q_1^2+q_2^2)\,\xi(0)} \int_0^\infty d\alpha_2$$
$$\cdot\;\exp\!\left[-\frac{\alpha_2}{2A_0}(p_2{}^2+M^2) - 2q_1\cdot q_2\,\xi(\alpha_2)\right] + \left(q_1 \leftrightarrow q_2\right) \tag{52}$$
$$G_{2,3}^{\mathrm{tr}}(p,p';q_1,q_2,q_3) = 2g^3\,A_0\,e^{-(q_1^2+q_2^2+q_3^2)\,\xi(0)} \sum_{\mathcal{P}(\{q\})} \int_0^\infty d\alpha_2\,d\alpha_3$$
$$\cdot\;\exp\!\left[-\frac{\alpha_2}{2A_0}(p_2{}^2+M^2) - \frac{\alpha_3}{2A_0}(p_3{}^2+M^2)\right] \tag{53}$$
$$\cdot\;\exp\!\left[-2q_1\cdot q_2\,\xi(\alpha_2) - 2q_2\cdot q_3\,\xi(\alpha_3) - 2q_1\cdot q_3\,\xi(\alpha_2+\alpha_3)\right]\,.$$

The expression for the $(2+1)$-point function in Eq. (52) may be written as
$$G_{2,1}^{\mathrm{tr}}(p,p';q) = 2g_{\mathrm{eff}}\,F(q^2)\;, \tag{54}$$
showing that the effective coupling constant for the meson-nucleon vertex is enhanced due to vertex corrections :
$$g_{\mathrm{eff}} = g\,A_0 > g\;. \tag{55}$$
This is qualitatively (but not quantitatively) similar to what one obtains from expanding the one-loop effective potential $V^{(1)}(\Phi,\varphi)$ (see II) up to the power of $\Phi^2\varphi$.

Not surprisingly the elastic meson-nucleon form factor

$$F(q^2) = e^{-\xi(0)q^2} \qquad (56)$$

is gaussian and defines the mean square radius of the dressed particle to be

$$\langle r^2 \rangle = 6\,\xi(0) = \frac{3}{\pi} \int_0^\infty dE \frac{1}{E^2} \left( \frac{1}{A(E)} - \frac{1}{A(0)} \right) . \qquad (57)$$

For the Feynman parametrization we obtain from Eq. (50)

$$\langle r^2 \rangle_F = \frac{3}{2} \frac{v^2 - w^2}{v^3} , \qquad (58)$$

which is a well-known result in the polaron literature [13]. Table 1 gives the numerical values obtained with the different parametrizations for the profile function. We also have included the perturbative result from the vertex correction in Fig. 4 b which reads

$$\langle r^2 \rangle_{\rm pert} = \frac{g^2}{4\pi^2} \int_0^1 du \frac{u^3}{[(1-u)m^2 + u^2 M^2]^2} . \qquad (59)$$

Table 1: Root-mean-square radius (in fm) of the dressed particle from Eq. (57). 'Feynman' signifies the result in the Feynman parametrization whereas 'improved' refers to the improved parametrization from Eq. (15). The radius calculated with the solution of the variational equations is denoted by 'variational'. For comparison the perturbative result is also shown.

| $\alpha$ | 'Feynman' | 'improved' | 'variational' | perturbative |
|---|---|---|---|---|
| 0.1 | 0.014 | 0.026 | 0.018 | 0.047 |
| 0.2 | 0.021 | 0.038 | 0.027 | 0.066 |
| 0.3 | 0.028 | 0.050 | 0.034 | 0.081 |
| 0.4 | 0.034 | 0.062 | 0.042 | 0.094 |
| 0.5 | 0.041 | 0.073 | 0.050 | 0.105 |
| 0.6 | 0.049 | 0.088 | 0.060 | 0.115 |
| 0.7 | 0.060 | 0.108 | 0.073 | 0.124 |
| 0.8 | 0.081 | 0.149 | 0.098 | 0.133 |

The values of the root-mean-square radius (Table 1) for the variational calculations differ much more from each other, as they do from the perturbative results, than the corresponding results for the residue in (II), even for small coupling. The reason for this difference at small coupling is that we have only calculated the $(2+1)$- (and higher) point function at zeroth order in $\langle \Delta S \rangle$. So, by construction, they are only 'forced' to agree at the tree level in perturbation theory (which they do). If one calculates them to first order in $\langle \Delta S \rangle$ the agreement for small couplings will be similar to the agreement for the residue of the two-point function determined in (II). Still, even at zeroth order the root-mean-square radii are quite consistent with each other. It should be noted that the radius is extremely sensitive to the behaviour of the profile function at small $E$, so the differences between the radii is a reflection of the differences between the profile functions seen in this region.

The radius (57) is due to the recoil of the nucleon when it absorbs the external meson while other mesons have been emitted into the meson 'cloud' surrounding the dressed particle. Since there is no direct meson-meson interaction the natural length scale for the radius therefore is the Compton wavelength $1/M = 0.21$ fm of the nucleon. The actual values are, of course, dependent on the strength of the coupling and, as seen in Table 1 the root-mean-square radius increases with $\alpha$. In principle for some large enough coupling it should become comparable to the empirical charge radius of the proton ( $\approx 0.8$ fm ), were it not for the instability which forces the coupling constant to be less than $\alpha_c$. This is probably an unreasonable comparison in any case – we are treating the (bare) nucleon as a point particle here, so the root-mean-square radius is only due to effects from the meson cloud.

## 4  Meson-Nucleon Elastic Scattering

We now turn to a more detailed discussion of the 'Compton' amplitude given by Eq. (53). It consists of two parts, the direct diagram and the crossed one shown in Figs. 5 a and 5 b where we have set $q \equiv -q_1$, $q' \equiv q_2$ for the meson and $p' \to -p'$ for the *outgoing* nucleon. In terms of the usual *Minkowski space* Mandelstam variables

$$s = -(q+p)^2, \quad t = -(p-p')^2, \tag{60}$$

we have for the crossed diagram

$$G_{2,2}^{\text{crossed}}(s,t) = 2g^2 A_0 \, e^{2m^2 \, \xi(0)} \int_0^\infty d\alpha_2 \, \exp\left[-\frac{\alpha_2}{2A_0}\left(s+t-M^2-2m^2\right) - (2m^2-t)\,\xi(\alpha_2)\right] \tag{61}$$

and for the direct diagram

$$G_{2,2}^{\text{direct}}(s,t) = 2g^2 A_0 \, e^{2m^2 \, \xi(0)} \int_0^\infty d\alpha_2 \, \exp\left[-\frac{\alpha_2}{2A_0}\left(-s+M^2\right) - (2m^2-t)\,\xi(\alpha_2)\right]. \tag{62}$$

It is worthwhile to point out some similarities and differences of these formulae with the usual skeleton expansion of higher Green functions: after the substitution $\tau = \alpha_2/(2A_0)$ in the proper-time integrals we indeed can identify the enhancement factor in front of the integrals as $G_{2,1}^2(q^2 = -m^2)$, i.e. the square of the meson-nucleon vertex function (52). However, the zeroth order propagator $G_{2,0}$ studied previously does *not* appear in Eqs. (61, 62) and consequently, none of its off-shell deficiencies will show up. This is because we only consider truncated Green functions which, by definition, are on the mass shell with respect to the external particles. The variational principle then yields well-behaved expressions even though the internal particles may be off-shell in a diagrammatic expansion.

When applying Eqs. (61, 62) to meson-nucleon scattering the only question which remains is the convergence of the integral representations for the 'Compton' amplitude. Whereas Eq. (61) for the crossed diagram converges for $s+t > M^2 + 2m^2$, which includes the physical region, this is not the case for the direct diagram. Indeed, Eq. (62) is only defined for $s < M^2$ and must be analytically continued for the physical values $s \geq (M+m)^2$.

### 4.1  Analytic Continuation

To perform this analytic continuation we must investigate the analytic properties of the function $\xi(\alpha_2)$ which has been defined in Eq. (45) as the Fourier cosine transform of the inverse profile function. Using the fact that the profile function is even in $E$ and assuming that it has the analytic structure shown in Fig. 2 we may again deform the integration contour and express $\xi(\alpha_2)$ by an integral running along both sides of the cut

$$\xi(\alpha_2) = \int_{E_0}^{\infty} dE\, \rho(E)\, e^{-E\alpha_2} \qquad (63)$$

where

$$\rho(E) = \frac{1}{2\pi E^2} \operatorname{Im} \frac{1}{A(iE+\epsilon)}\, . \qquad (64)$$

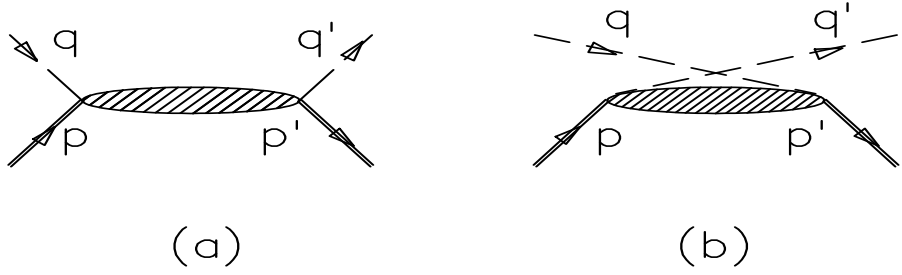

Figure 5: 'Compton' amplitude for meson-nucleon scattering in the zeroth order variational calculation: (a) direct diagram, (b) crossed diagram.

The explicit form of the weight function $\rho(E)$ for the various parametrizations can be found in the Appendix. In deriving Eq. (63) we have tacitly assumed that the profile function $A(E)$ has *no zeroes* in the upper plane which can be verified for the 'improved' parametrization. Of course, if this is not the case one has to include the contribution from the corresponding poles in the integrand for $\xi(\alpha_2)$.

The integral representation (63) can now be used to evaluate the $\alpha_2$-integral for the direct diagram. After expanding the exponential and integrating term by term we obtain

$$G_{2,2}^{\text{direct}}(s,t) = 2g^2 A_0\, e^{2m^2\, \xi(0)} \left[ \frac{2A_0}{M^2-s} + \sum_{n=1}^{\infty} \frac{(t-2m^2)^n}{n!} \int_{E_0}^{\infty} dE_1\ldots dE_n \right.$$
$$\left. \cdot\, \rho(E_1)\ldots \rho(E_n)\, \frac{1}{\frac{1}{2A_0}(M^2-s) + E_1 + \ldots E_n} \right] . \qquad (65)$$

Although Eq. (65) has been derived for $s < M^2$, it now can be used to define the direct amplitude for other values as well, in particular on the upper side of the $s$-cut ($s \to s + i\epsilon$) which corresponds to the physical region. The infinite series (65) may then be resummed by means of the representation

$$\frac{1}{a - i\epsilon} = i \int_0^{\infty} d\tau\, e^{-i\tau(a-i\epsilon)}\, . \qquad (66)$$

This gives

$$G_{2,2}^{\text{direct}}(s,t) = 2g^2 A_0\, e^{2m^2\, \xi(0)}\, i \int_0^{\infty} d\tau\, \exp\left[ -i\frac{\tau}{2A_0}(-s+M^2) - (2m^2-t)\,\xi(i\tau) \right] . \qquad (67)$$

Of course, Eq. (67) could have been derived more easily from Eq. (62) by the simple replacement

$$\alpha_2 = i\tau, \qquad (68)$$

which amounts to the usual Wick rotation back to Minkowski proper time. However, the derivation makes clear the crucial role played by the analytic properties of the profile function $A(E)$ which determines the function $\xi(\tau)$. In addition, the expanded version (65) of the direct amplitude is very useful to study thresholds and the threshold behaviour of the imaginary part of the scattering amplitude. This will be done in the next subsection.

## 4.2 Threshold Behaviour and Optical Theorem

As the zeroth order residue is one, the truncated Green function equals the scattering amplitude

$$\mathcal{T}(s,t) = G_{2,2}^{\text{tr}}(s,t). \qquad (69)$$

This may be verified by calculating the Born terms in perturbation theory. Using, for example, the conventions for the Feynman rules advocated by Muta [15] [2] one obtains

$$\mathcal{T}^{\text{Born}} = (2g)^2 \left[ \frac{1}{M^2 + (p+q)^2} + \frac{1}{M^2 + (p-q')^2} \right]. \qquad (70)$$

which is identical with Eq. (43).

The differential cross section in the center-of-mass system can be expressed as (Ref. [14], p. 245)

$$\frac{d\sigma}{d\Omega} = \frac{1}{64\pi^2 s} |\mathcal{T}(s,t)|^2 \qquad (71)$$

and the total cross section by means of the optical theorem as

$$\sigma_{\text{tot}}(s) = \frac{1}{2|\mathbf{p}|\sqrt{s}} \operatorname{Im} \mathcal{T}(s, t=0). \qquad (72)$$

Here $\mathbf{p}$ is the center-of-mass momentum. The Born amplitude (70) is, of course, purely real thereby leading to a vanishing total cross section and a violation of the optical theorem. The direct amplitude in our zeroth order variational calculation, however, develops an imaginary part after the analytic continuation has been made. Indeed, using the expanded version (65) of the direct amplitude with $s \to s + i\epsilon$ we have (in the physical region)

$$\operatorname{Im} \mathcal{T}(s,0) = 2\pi g^2 A_0 \, e^{2m^2 \xi(0)} \sum_{n=1}^{\infty} \frac{(-2m^2)^n}{n!} \int_{E_0}^{\infty} dE_1 \ldots dE_n \, \rho(E_1) \ldots \rho(E_n)$$
$$\cdot \, \delta\left( \frac{1}{2A_0}(M^2 - s) + E_1 + \ldots E_n \right). \qquad (73)$$

The $\delta$-function may be used to perform, say, the $E_1$-integration which gives a nonzero result only if the singularity lies within the integration interval. Since all integrations over the $E_i$'s start at $E_0$ where the branch point of the profile function is located, we find that the $n$ - th term in Eq. (73) only makes a contribution to the imaginary part of the scattering amplitude if

$$s \geq s_{\text{thresh}}^{(n)} = M^2 + 2nA_0 E_0 \qquad n = 1, 2 \ldots \qquad (74)$$

---

[2]These conventions produce the minimal number of $i$'s, do not change the sign of the propagator when a Wick rotation is made and directly yield the transition matrix element. Our Lagrangian (1) gives a factor $2g$ at each vertex.

In particular, between the elastic and the first inelastic threshold for meson-nucleon scattering only *one* term in the sum (73) contributes and we have

$$\operatorname{Im} \mathcal{T}(s,0) \;=\; -4\pi g^2 m^2 A_0 \, e^{2m^2 \, \xi(0)} \, \rho\left(\frac{s - M^2}{2A_0}\right) , \qquad s^{(1)}_{\mathrm{thresh}} \leq s \leq s^{(2)}_{\mathrm{thresh}} . \tag{75}$$

We see that the cut structure of the profile function $A(E)$ not only determines the threshold position (74) for meson-nucleon scattering but also directly the total cross section.

Let us now discuss qualitatively the effects produced by the different parametrizations which we have used in the variational calculation on the pole of the 2-point function. We immediately recognize that the rather small deviations which these parametrizations have given for the self-energy lead to very large ones for the scattering cross section. For example, the Feynman parametrization has no cuts at all but only poles and zeroes in the profile function. Consequently there is only a very pathological (and unphysical) scattering cross section when this parametrization is continued to Minkowski space: from Eq. (A.2) in the Appendix we see that the total cross section would be a sum of $\delta$-functions !

The 'improved' parametrization fares better at first sight: it has a cut in the upper $E$-plane starting at $E = iE_0 = iw$ . Therefore

$$\sqrt{s^{(1)}_{\mathrm{thresh}}} - M \;=\; \sqrt{M^2 + 2A_0 E_0} \;-\; M \tag{76}$$

which should be equal to the meson mass $m$. Using Table 2 in (II) the right-hand side of Eq. (76) is found to vary between 408 MeV at $\alpha = 0.1$ and 275 MeV at $\alpha = 0.8$ . Although this is a far cry from the correct value 140 MeV it could be easily cured by performing the minimization with the constraint that the correct threshold is reproduced. A more serious problem, however, is that near threshold the weight function $\rho(E)$ behaves like (see Eqs. (A.5) and (A.10) )

$$\rho(E) \;\stackrel{E \to w}{\longrightarrow}\; C_I \, (E - w) \tag{77}$$

with a *positive* constant $C_I$. In this kinematical region the center-of-mass energy is

$$s \;=\; \left( \sqrt{M^2 + \mathbf{p}^2} \;+\; \sqrt{m^2 + \mathbf{p}^2} \right)^2 \;\stackrel{\mathbf{p} \to 0}{\longrightarrow}\; s^{(1)}_{\mathrm{thresh}} \;+\; \mathbf{p}^2 \, \frac{(M + m)^2}{Mm} . \tag{78}$$

Using the threshold condition (76) we thus find that the weight function in Eq. (75) behaves like

$$\rho\left(\frac{s - M^2}{2A_0}\right) \;\stackrel{\mathbf{p} \to 0}{\longrightarrow}\; C_I \, \frac{\mathbf{p}^2}{2A_0} \, \frac{(M+m)^2}{Mm} . \tag{79}$$

This has two disastrous consequences: firstly, the imaginary part of the amplitude vanishes like $\mathbf{p}^2$ (i.e. not like $|\mathbf{p}|$) leading to a zero total cross section (72) at threshold and, secondly, away from threshold $\sigma_{\mathrm{tot}}$ becomes *negative* when calculated by means of the optical theorem !

We will see in the next subsection that both these failures are not due to the zeroth order variational approximation but are a consequence of an inadequate form of the retardation function.

## 4.3 An Extended Parametrization of the Retardation Function

It is not surprising that many of the parametrizations for the retardation function investigated in (I, II) turn out to be inadequate for describing meson nucleon scattering. By means of a simple approximation we have found in (I) that the variational principle for the self-energy is sensitive mostly to *small* values of the proper time difference $\sigma$. On the other hand, a scattering amplitude near threshold probes *large* values of this variable as can be inferred, for example, from Eq. (67).

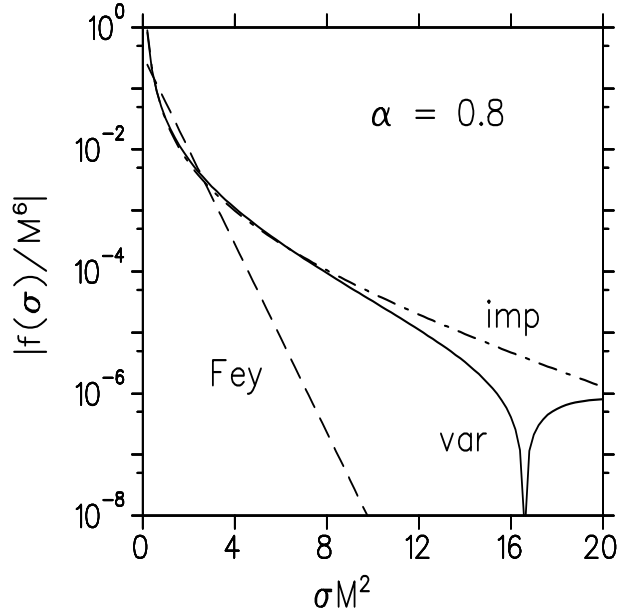

Figure 6: Retardation functions $f(\sigma)$ versus proper time $\sigma$ for various parametrizations: 'Fey' denotes Feynman's parametrization, 'imp' the 'improved' parametrization and 'var' the solution of the variational equations. Note that for large $\sigma$ the variational retardation function becomes negative.

Fortunately the variational principle tells us what the *best* form for the retardation function is for all values of $\sigma$, including $\sigma \to \infty$. Indeed, performing an integration by parts in Eq. (16) one obtains

$$f_{\rm var}(\sigma) = \frac{g^2}{32\pi^2} \frac{1}{\mu^4(\sigma)} \left[ \exp\left(-\frac{\lambda^2 M^2 \sigma^2}{2\mu^2(\sigma)}\right) - \frac{m^2}{2}\mu^2(\sigma) \int_0^1 du\, e\left(m\mu(\sigma), \frac{\lambda M \sigma}{\mu(\sigma)}, u\right) \right]. \qquad (80)$$

Replacing $\mu^2(\sigma)/\sigma$ by some average value one sees that the 'improved' parametrization just corresponds to the first term in Eq. (80). However, the second term dominates at large $\sigma$ and makes the retardation function even *negative*. This is shown in Fig. 6 where the retardation functions investigated in (II) are plotted as function of $\sigma$. It can also be demonstrated analytically in the following way: recalling the asymptotic limit of the pseudotime given in Eq. (20) we evaluate the $u$-integral by Laplace's method

$$\int_0^1 du\, e^{-E(u)\sigma} \stackrel{\sigma \to \infty}{\longrightarrow} \sqrt{\frac{2\pi}{\sigma E''(u_0)}} e^{-E(u_0)\sigma}. \qquad (81)$$

Here $u_0$ is the value which minimizes the function $E(u)$ in the interval between 0 and 1. For the case at hand (see Eq. (17) )

$$E(u) = \frac{m^2}{2A_0}\left(\frac{1}{u} - 1\right) + \frac{A_0 \lambda^2 M^2}{2} u, \qquad (82)$$

so that

$$E\left(u_0 = \frac{m}{A_0 \lambda M}\right) = \lambda M m - \frac{m^2}{2A_0}, \qquad E''(u_0) = \frac{A_0^2 \lambda^3 M^3}{m} > 0. \qquad (83)$$

Since $E(u_0)$ is smaller than the value $\lambda^2 M^2 A_0/2$ which governs the exponential fall-off of the first term, we see that the variational retardation function asymptotically behaves like

$$f_{\rm var}(\sigma) \stackrel{\sigma \to \infty}{\longrightarrow} -\frac{g^2}{32\pi^2} \sqrt{\frac{\pi}{2}} \left(\frac{m}{\lambda M}\right)^3 \frac{m}{\sigma^{3/2}} e^{-E(u_0)\sigma}. \qquad (84)$$

Evidence for a sign change of the variational retardation function for large $\sigma$ can already be found in the 'dip' of $A_{\rm var}(E)$ near $E = 0$ which has been observed in the plots given in (II). As a side remark we note that the threshold for meson-nucleon scattering determined from Eqs. (84) and (83) is now much closer to the physical value than with the 'improved' parametrization: with $E_0 = E(u_0)$ and Table 3 of Ref. (II) the right-hand side of Eq. (76) now varies between 120 MeV at $\alpha = 0.1$ and 99 MeV at $\alpha = 0.8$. Far more important than the remaining discrepancies (which may be easily removed by a constraint calculation) is the sign of the asymptotic retardation function and the fractional power of $\sigma$ which accompanies the exponential decay. This is because a retardation function of the asymptotic form

$$f(\sigma) \stackrel{\beta \to \infty}{\longrightarrow} \frac{C}{\sigma^\gamma} e^{-E_0 \sigma} \tag{85}$$

leads to the following behaviour of the weight function near the branch point $E_0$

$$\rho(E) \propto (E - E_0)^{\gamma - 1}, \tag{86}$$

the proportionality constant having the same sign as the constant $C$ in Eq. (85). This is demonstrated in the Appendix. Thus the asymptotic behaviour (84) "kills two birds with one stone" : it produces a positive total cross section and the correct threshold behaviour of the imaginary part of the scattering amplitude to make the total cross section finite at $\mathbf{p} = 0$. In addition, it leads to a lower minimum of the variational functional thereby showing that such a form is a consequence of the variational principle applied to the present model and not an arbitrary parametrization.

We incorporate the behaviour of the variational retardation function at small and large $\sigma$ by the following ansatz for an 'extended' parametrization

$$f_E(\sigma) = \frac{C_1}{\sigma^2} \left[ e^{-w_1 \sigma} - C_2 \sqrt{\sigma}\, e^{-w_2 \sigma} \right], \tag{87}$$

and require $w_1 > w_2$ and $C_2 > 0$. The associated profile function is

$$A_E(E) = 1 + \frac{2C_1}{E^2} \left[ 2E \arctan \frac{E}{w_1} - w_1 \ln \left( 1 + \frac{E^2}{w_1^2} \right) \right.$$
$$\left. - 4 C_2 \sqrt{\pi} \left( \sqrt{\frac{1}{2}\left(\sqrt{w_2^2 + E^2} + w_2\right)} - \sqrt{w_2} \right) \right], \tag{88}$$

having branch points at $E = iE_0 = iw_2$ and $E = iE_0' = iw_1$. In the Appendix the explicit expressions for the discontinuity across the cut and the weight function $\rho_E(E)$ are listed. The value of the 'extended' profile function at $E = 0$ is

$$A_E(0) = 1 + 2C_1 \left[ \frac{1}{w_1} - C_2 \frac{\sqrt{\pi}}{2 w_2^{3/2}} \right]. \tag{89}$$

We determine the variational parameters $C_i, w_i, i = 1, 2$ under the constraint that the elastic threshold for meson-nucleon scattering is at $s_{\rm thresh}^{(1)} = (M + m)^2$. According to Eq. (74) this requires that the relation

$$A_0 E_0 = Mm + \frac{m^2}{2} \tag{90}$$

with $E_0 = w_2$ should be fulfilled during variation.

As the imaginary part of the forward scattering amplitude below the first inelastic threshold is fully determined by the weight function ( see Eq. (75) ) we can now derive the total cross section in analytic form, in particular at threshold. After some algebra we obtain

$$\sigma_{\rm tot}(|\mathbf{p}| = 0) = 16\pi^2 \alpha \sqrt{\frac{A_0}{2\pi Mm}} \frac{C_1 C_2 M^2 m^2}{w_2^4} \frac{e^{2m^2 \xi(0)}}{R^2(w_2)} \tag{91}$$

where $R(w_2)$ is the real part of the profile function at the branch point (see the Appendix). Eq. (91) shows that the constant total cross section at threshold is entirely due to the new term $C_2$ in the 'extended' parametrization. Since we expect the overall strength $C_1$ to be of order $g^2$ (see Eq. (80) ) the expression (91) actually is of order $g^4$ in the perturbative limit, in agreement with the square of the Born amplitude (70). This is a necessary condition for fulfilling the optical theorem and maintaining unitarity in our variational approach.

## 5 Numerical Results and Discussion

We first turn to the minimization of the variational functional on the pole of the propagator

$$ - M_1^2 \leq (\lambda^2 - 2\lambda) M^2 + 2 (\Omega + V) \tag{92}$$

for the 'extended' parametrization (87). This minimization is done with respect to the 'velocity' parameter $\lambda$ and the variational parameters which enter the profile function $A(E)$. The explicit expressions for the quantities $\Omega$ and $V$, which are functionals of the profile function and the pseudo-time, can be found in (II), where also the numerical procedures are described. Again we have chosen $M = 939$ MeV and $m = 140$ MeV for the masses. For the present calculation the threshold constraint (90) was used to eliminate the parameter $C_2$ via Eq. (89) so that a 4-parameter minimization of Eq. (92) had to be performed.

Table 2 gives the parameters of the 'extended' parametrization obtained in this way as well as some quantities of interest derived from them. In view of the expressions (80) and (84) for the variational retardation function we write the strength parameters as

$$ C_1 = x_1 \frac{g^2}{32\pi^2} = x_1 \frac{\alpha}{8\pi} M^2 \tag{93}$$

$$ C_2 = x_2 \, m \sqrt{\frac{\pi}{2} \left(\frac{m}{M}\right)^3} \tag{94}$$

and list the dimensionless numbers $x_1, x_2$. Whereas $x_1$ stays close to its perturbative value one, the strength parameter $x_2$ for the asymptotic term in the retardation function turns out to be much larger even at small coupling and changes rapidly when the coupling constant $\alpha$ is increased.

Comparing the numerical values of the intermediate mass $M_1$ with the ones obtained in (II) we see that the 'extended' parametrization is slightly better than the 'improved' one but, of course, inferior to the (unconstrained) variational solution. This is more visible in Fig. 7 where

$$ \Delta M_1^2 = M_1^2 - M_1^2 \Big|_{\text{improved}} \tag{95}$$

is plotted as a function of the coupling constant $\alpha$. For comparison we also have included the perturbative result

$$ M_1^2 \Big|_{\text{pert}} = M^2 - \frac{g^2}{4\pi^2} \int_0^1 du \, \ln \left( 1 + \frac{M^2}{m^2} \frac{u^2}{1-u} \right) = M^2 \, ( 1 - 1.0214 \, \alpha ) \, . \tag{96}$$

As expected the perturbative result and the minimal value with Feynman's parametrization always lie above the value obtained with the 'improved' parametrization. Note that at larger coupling constants one has to scale these results by several orders of magnitude in order to display them in the graph. In contrast, the gain achieved with the 'extended' parametrization and the variational calculation is rather modest. As mentioned before this is due to the sensitivity of the self-energy to small values of $\sigma$ so that the completely different asymptotic behaviour of the retardation function which is built into the 'extended' parametrization is not fully reflected in the value of $M_1^2$. This is, of course, well known from applications of the Ritz variational principle in quantum mechanics : even very refined wave functions lower the ground state energy only by a small amount compared to crude ones. However, calculation of other observables may lead to very different results. In the present case this phenomenon is amplified by the need of an analytic continuation to Minkowski space in which scattering takes place.

Table 2: Variational calculation for the nucleon self-energy in the Wick-Cutkosky model using the 'extended' parametrization (87) for the retardation function. The parameters $x_i$, $w_i$, $i = 1, 2$ obtained from minimizing Eq. (92) are given as well as $\lambda$ for different values of the coupling constant $\alpha$. They are constrained such that the correct elastic threshold is obtained. The dimensionless strength parameters $x_1$ and $x_2$ are defined in Eqs. (93) and (94). The lower part of the table lists the mass $M_1$, the value of the profile function at $E = 0$, the first order residue (see (II)) and the root-mean-square radius (57) of the dressed particle.

|  | $\alpha = 0.1$ | $\alpha = 0.2$ | $\alpha = 0.3$ | $\alpha = 0.4$ | $\alpha = 0.5$ | $\alpha = 0.6$ | $\alpha = 0.7$ | $\alpha = 0.8$ |
|---|---|---|---|---|---|---|---|---|
| $x_1$ | 1.0093 | 1.0182 | 1.0296 | 1.0426 | 1.0600 | 1.0863 | 1.1266 | 1.1905 |
| $x_2$ | 2.189 | 2.564 | 2.936 | 3.682 | 4.803 | 6.363 | 7.246 | 4.550 |
| $\sqrt{w_1}$ [MeV] | 632.7 | 609.0 | 584.8 | 554.0 | 518.0 | 476.6 | 437.5 | 399.5 |
| $\sqrt{w_2}$ [MeV] | 373.0 | 370.0 | 366.5 | 362.5 | 357.8 | 351.8 | 343.1 | 324.8 |
| $\lambda$ | 0.97297 | 0.94389 | 0.91223 | 0.87718 | 0.83739 | 0.79033 | 0.72987 | 0.62457 |
| $M_1$ [MeV] | 890.25 | 839.78 | 787.43 | 732.97 | 676.20 | 616.98 | 555.57 | 493.45 |
| $A(0)$ | 1.0150 | 1.0321 | 1.0518 | 1.0751 | 1.1037 | 1.1415 | 1.2002 | 1.3393 |
| $Z^{(1)}$ | 0.96088 | 0.91919 | 0.87429 | 0.82523 | 0.77042 | 0.70679 | 0.62687 | 0.49289 |
| $\langle r^2 \rangle^{1/2}$ [fm] | 0.017 | 0.025 | 0.032 | 0.038 | 0.043 | 0.052 | 0.076 | 0.135 |

With the variational parameters of the 'extended' calculation fixed we can now calculate the *imaginary* part of the scattering amplitude very easily from Eq. (75). The first column of Table 3 gives the total cross section (72) at $\mathbf{p} = 0$ obtained from the optical theorem for various coupling constants and Fig. 8 displays $\sigma_{\text{tot}}(|\mathbf{p}|)$ for two different couplings below the first inelastic threshold. It should be noted that the threshold condition (74) associated with the first branch point $E_0 = w_2$ automatically gives the correct higher (inelastic) thresholds. However, there are additional higher thresholds coming from the second branch point $E_0' = w_1$ in the profile function of the 'extended' parametrization. Since $w_1$ is decreasing for larger coupling constants (see Table 2) these additional thresholds may even come to lie below the first inelastic threshold ( at $|\mathbf{p}| = 214.2$ MeV ) which is seen, e.g., as a cusp in the total cross section for $\alpha = 0.6$ near $|\mathbf{p}| = 180$ MeV. It is, of course, possible to fix these additional thresholds at the correct physical values in a similar way as was done for the elastic threshold. We will not do this in the present work but concentrate on the kinematical region close to the elastic threshold.

It is much more demanding to calculate the *real* part of the scattering amplitude at threshold in a reliable way. Numerical problems do not arise, of course, in the crossed amplitude which has the euclidean proper time integral representation (61) but in the evaluation of the oscillatory integrals of the direct amplitude (67). The expansion (65) is of no help since *all* powers of $\rho(E)$ contribute to the real part in a given interval between thresholds. In addition, one would have to perform high dimensional principal value integrals numerically which is not a very promising procedure. Rather, we have employed the subroutines DQDAWF from the IMSL Mathematical Library and D01ASF from the NAG Fortran Library to perform the required sine and cosine transforms (over an infinite range) numerically. The latter routine was used to calculate $\xi(i\tau)$ with high accuracy whereas the former one served to evaluate the $\tau$-integral in (67) with less accuracy. To obtain a properly convergent integral the corresponding Born term first had to be subtracted, i.e. the decomposition

$$e^{-(2m^2-t)\,\xi(i\tau)} \;=\; 1 + \left[\, e^{-(2m^2-t)\,\xi(i\tau)} \;-\; 1 \,\right] \tag{97}$$

was made in the integrand. While the first term generates the Born term for the direct diagram the last one is now amenable to numerical integration as $\xi(i\tau)$ goes to zero for large values of $\tau$ (see Eq. (A.27) ). We have checked the numerical stability of our program by treating the crossed diagram in the same fashion and comparing the results with the euclidean proper time representation (61). Table 3 also contains the values for

$$\sigma_{\rm el}\,(|{\bf p}| = 0) \;=\; \frac{1}{16\pi s}\left[\, {\rm Re}\;\; {\cal T}\,(s, t=0) \,\right]^2 \tag{98}$$

at threshold where $s = (M+m)^2$ and the imaginary part of the scattering amplitude vanishes. For comparison we also list the corresponding Born cross section (see Eq. (70) )

$$\sigma_{\rm el}^{mboxBorn}\,(|{\bf p}|=0) \;=\; \frac{4\pi\alpha^2}{(M+m)^2\,\left(1 - \frac{m^2}{4M^2}\right)} \;=\; 0.4226\,\alpha^2\;\;[\,{\rm fm}^2\,]. \tag{99}$$

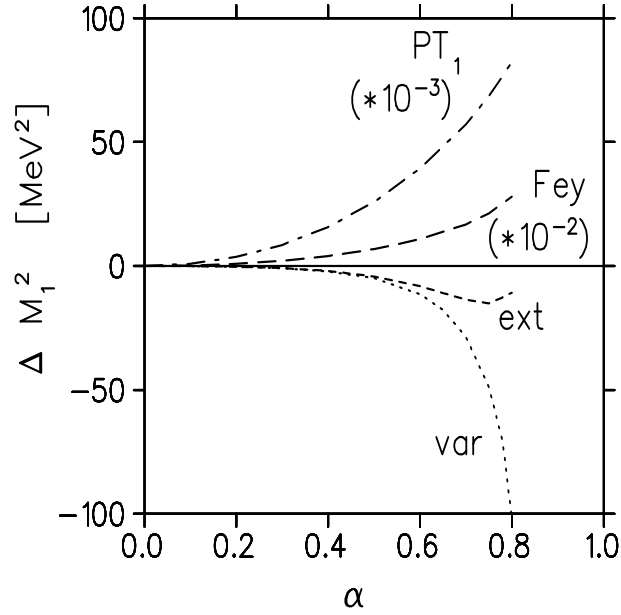

Figure 7: Minimum of the variational functional (92) relative to the 'improved' parametrization as a function of the coupling constant for various parametrizations. The notation is as in Fig. 6. In addition, 'PT$_1$' denotes the result from first-order perturbation theory and 'ext' the one from the 'extended' parameterization (87).

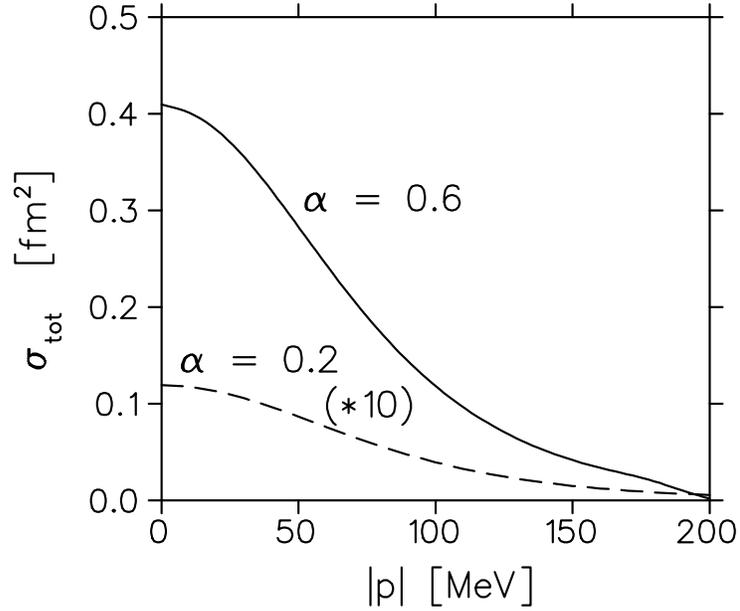

Figure 8: Total cross section for two coupling constants as a function of the center-of-mass momentum $|\mathbf{p}|$. The values for $\alpha = 0.2$ have been multiplied by a factor 10. The 'extended' parametrization of the retardation function has been employed.

Table 3: Total and elastic meson-nucleon cross sections at threshold in the 'extended' parametrization for different values of the coupling constant $\alpha$. The total cross section (91) has been evaluated assuming the optical theorem whereas the elastic cross section has been obtained from the square of the scattering amplitude at threshold. For comparison the elastic Born cross section (99) is also given. The last column lists the 'unitarity ratio' between elastic and total cross section.

| $\alpha$ | $\sigma_{\rm tot}$ [fm$^2$] | $\sigma_{\rm el}$ [fm$^2$] | $\sigma_{\rm el}^{\rm Born}$ [fm$^2$] | $\sigma_{\rm el}/\sigma_{\rm tot}$ |
|---|---|---|---|---|
| 0.1 | 0.0024 | 0.0045 | 0.0042 | 1.889 |
| 0.2 | 0.0119 | 0.0195 | 0.0169 | 1.634 |
| 0.3 | 0.0330 | 0.0478 | 0.0380 | 1.448 |
| 0.4 | 0.0802 | 0.0942 | 0.0676 | 1.176 |
| 0.5 | 0.183 | 0.168 | 0.106 | 0.916 |
| 0.6 | 0.409 | 0.288 | 0.152 | 0.704 |
| 0.65 | 0.585 | 0.375 | 0.179 | 0.640 |
| 0.7 | 0.779 | 0.495 | 0.207 | 0.636 |
| 0.75 | 0.929 | 0.665 | 0.238 | 0.716 |
| 0.8 | 0.835 | 0.956 | 0.270 | 1.145 |

We observe that at larger coupling constant the Born cross section is enhanced by up to a factor 3 due to final state interactions, vertex corrections and self-energy effects which are all (approximately) contained in our result. Most important are the vertex corrections because the enhancement can be explained, with an accuracy of better than 10 % , by replacing the coupling constant $g$ which enters

Eq. (99) by the effective coupling constant $g_{\text{eff}}$. More precisely but numerically nearly identical is the replacement $2g \to G_{2,1}^{\text{tr}}(q^2 = -m^2)$ (see Eq. (54)) which leads to

$$\sigma_{\text{el}}(|\mathbf{p}| = 0) \approx \frac{4\pi\alpha^2 A_0^4}{(M+m)^2 \left(1 - \frac{m^2}{4M^2}\right)} e^{4m^2 \xi(0)} . \tag{100}$$

Since the momentum transfer is small compared to the nucleon mass the form factor stays very close to one and the main enhancement effect comes from the factor $A_0^4$, i.e. the effective coupling constant.

The 'unitarity ratio'

$$U_0 = \frac{\sigma_{\text{el}}(|\mathbf{p}| = 0)}{\sigma_{\text{tot}}(|\mathbf{p}| = 0)} \tag{101}$$

decreases from 1.9 to 0.6 when the coupling is varied from $\alpha = 0.1$ to $\alpha = 0.7$ before rising again. These values can be understood semi-quantitatively in the following way: we first rewrite Eq. (91) as

$$\sigma_{\text{tot}}(|\mathbf{p}| = 0) = \pi\alpha^2 A_0^{9/2} \frac{M^2}{(M+\frac{m}{2})^4} \frac{x_1 x_2}{R^2(w_2)} e^{2m^2 \xi(0)} \tag{102}$$

where we have used the definitions (93, 94) for the strength parameters. We then assume that the elastic cross section can be approximated by Eq. (100) which gives for the ratio

$$U_0 \approx \frac{4}{x_2} \cdot \frac{\left(1 + \frac{m}{2M}\right)^4}{\left(1 + \frac{m}{M}\right)^2 \left(1 - \frac{m^2}{4M^2}\right)} \cdot e^{2m^2 \xi(0)} \cdot \frac{R^2(w_2)}{x_1 \sqrt{A_0}} . \tag{103}$$

Note that the large enhancement due to fourth power of $A_0$ has cancelled. In addition, the second factor is $1 + \mathcal{O}(m^2/M^2)$, the form factor is practically one and the last factor also turns out to be very close to one (except at $\alpha = 0.8$ where it is 1.17 ). Thus for nearly all accessible coupling constants one has the simple result

$$U_0 \approx \frac{4}{x_2} . \tag{104}$$

From Table 2 we see that the dimensionless parameter $x_2$ grows from 2.2 at small coupling to over 7 at $\alpha = 0.7$ before declining again and that the approximation (104) accounts rather well for the values $\sigma_{\text{el}}/\sigma_{\text{tot}}$ listed in Table 3.

The reverse procedure also works satisfactorily as can be seen in Fig. 9 : Here we have plotted the unitarity ratio for the 'extended' parametrization together with the results from a variational calculation in which $x_2$ has been fixed to the value $x_2 = 4$. (This leads to a minimal value of the variational functional which is nearly as good as the one from the unconstrained 'extended' parametrization.) Except for coupling constants close to the critical coupling we now observe equality of elastic and total cross section to a much better degree. It is clear that a fine tuning of the parameter $x_2$ could lead to a completely unitary result, at least at $|\mathbf{p}| = 0$ .

However, our aim is *not* to unitarize the scattering amplitude but to evaluate the imaginary part of the amplitude as a prediction of our variational approach. To what extent unitarity is fulfilled [3] thus serves as a severe test of our approximation scheme, in particular near threshold. In contrast, imposing unitarity, e.g. by considering the Born terms as K-matrix elements [16], is an ad-hoc procedure which is applied on top of an approximate calculation which violates unitarity to a much larger extent.

We also have performed a partial-wave projection of the scattering amplitude. Given the particular t-dependence of Eqs. (61) and (67) this can be done analytically. Fig. 10 shows the Argand diagram for the s-wave at $\alpha = 0.5$. We observe that up to $|\mathbf{p}| \simeq 60$ MeV/c the s-wave amplitude remains on the unitarity circle before appreciable deviations occur.

---

[3]Of course, here we disregard the instabiltity of the Wick-Cutkosky model and assume that the theory is unitary below the critical coupling.

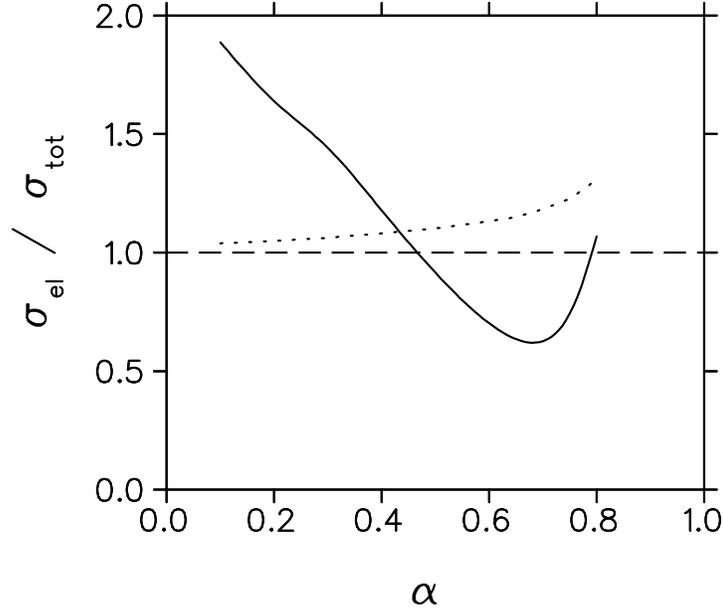

Figure 9: Ratio of elastic to total cross section at threshold as a function of the coupling constant $\alpha$. The full line gives the result using the 'extended' parametrization from Table 2, whereas the dotted line follows from a variational calculation in which the strength parameter $x_2 = 4$ has been kept fixed.

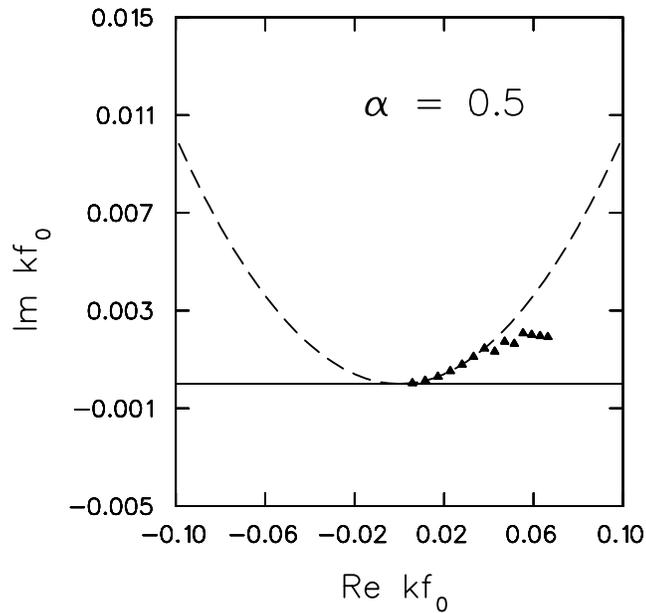

Figure 10: Argand plot for the s-wave scattering amplitude with the 'extended' parametrization of the retardation function at $\alpha = 0.5$. The triangles denote the values for different center-of-mass momenta from 0 to 140 MeV/c in intervals of 10 MeV/c.

# 6 Summary and Conclusion

We have extended the polaron variational approach in the scalar Wick-Cutkosky model to the scattering or absorption of an arbitrary number of mesons from the dressed nucleon. The retarded quadratic trial action whose parameters have been fixed variationally on the pole of the nucleon propagator was employed instead of the exact effective action. This constitutes the zeroth order approximation in a systematic expansion of the Green functions around the trial action and is similar to a calculation in quantum mechanics when the variational wavefunction determined from minimizing the energy is used to calculate other observables.

We have seen that already this lowest order gives sensible results: although only agreement with the tree level calculations is assured in the perturbative limit, the variation of the parameters effectively sums up parts of higher diagrams up to all orders. A nice example of this is the $(2+n)$-point function in Section 3. This expression contains pieces which may be identified with Feynman diagrams of arbitrary complexity and to any order in the coupling. A look at the explicit expressions also shows that the zeroth order variational approximation *exponentiates* lowest order results in a particular way. This is a welcome feature since exponentiation is a frequently used recipe to extend the range of validity of perturbation theory. As a caveat one should add, however, that the off-shell behaviour of the zeroth order propagator was found to be unsatisfactory if the on-shell variational parameters are used to extrapolate away from the nucleon pole. No problems arise if only truncated (on-shell) Green functions are considered as done in this work.

Independent of the (not very realistic) model field theory which we consider here the general expressions even may be used in phenomenological applications by parametrizing the functions $\xi(\tau)$ or $\rho(E)$ which fully determine all on-shell Green functions. Once this is done for elastic meson-nucleon scattering all other multi-meson processes could be predicted.

In the present work we first have studied in some detail the vertex function for the absorption of a virtual meson on the dressed nucleon. Given that the trial action is quadratic in the nucleon trajectories it is not surprising that the corresponding form factor turned out to be gaussian. For the radius of the dressed particle we obtained a similar expression as in the polaron case. Since there is no tree-level radius the numerical results showed some differences between the various parametrizations which enter the trial action.

We then concentrated on the zeroth order 'Compton' amplitude for meson-nucleon elastic scattering which has a much richer physical content. This required an analytic continuation of the variational results obtained in euclidean space back into Minkowski space. We have shown that the key to a successful description of scattering at threshold is the proper form of the retardation function $f(\sigma)$ which multiplies the quadratic trial action. For example, Feynman's classic parametrization used for the polaron, also employed by Mano [8] for the nucleon self-energy, is ruled out as it gives rise to a totally inappropriate analytic structure in the complex energy plane. Incidentally Mano himself writes in the conclusion of his work: "We also note that this method can be extended, though the accuracy of the result may be not very high, to another problem such as the scattering of the meson by the nucleon by using the best estimate of the real action."

We already had found in previous work (I, II) that a good variational calculation of the self-energy requires an $1/\sigma^2$-singularity for small $\sigma$. We now find in addition that scattering near the elastic threshold demands a specific behaviour of the retardation function for asymptotic values of the proper time $\sigma$. Remarkably the variational solution for the retardation function which was derived on the nucleon pole already contains that information (see Eq. (84) ) and has guided us to the appropriate form of the retardation function in both limits. We have incorporated the small- and large-$\sigma$ behaviour in an 'extended' parametrization that gives the correct elastic threshold and an imaginary part of the scattering amplitude which grows linearly with the center-of-mass momentum away from threshold.

Such a momentum-dependence leads to the expected constant total cross section at threshold. In addition, the 'extended' parametrization gives a lower value of the variational functional than the previously studied parametrizations (see Fig. 7). As our variational principle is a *minimum* principle, this minimum value is a clear measure of the quality of the corresponding ansatz.

By means of the optical theorem we have calculated the total cross section and compared it with the integrated elastic cross section. The latter shows a considerable enhancement over the Born cross section at larger coupling. This is mainly due to vertex corrections which give rise to a larger effective coupling constant. At threshold we found a ratio of elastic to total cross section between 1.9 and 0.6 depending on the coupling constant. By a simple analytic approximation we were able to show that this ratio is mainly determined by the strength of the asymptotic part of the retardation function and that a slight readjustement gives nearly unitary results. It is remarkable that threshold position, threshold behaviour of the total cross section and unitarity basically can fix *all* quantities in the asymptotic form (85) of the retardation function.

Although unitarity requires strict equality of elastic and total cross section below the first inelastic threshold (if the instability of the model is disregarded), we still consider the numerical result satisfactory in several respects. First, it is a prediction of our zeroth order variational principle without invoking any unitarizing procedure. Second, the 'extended' parametrization is still not the optimal variational solution as Fig. 7 shows. Given the sensitivity of the analytic continuation procedure to small changes in the retardation function one may expect a further improvement of the unitarity ratio when more refined ansätze are used. A solution of the variational equations with the correct elastic threshold as constraint would be the optimal procedure if the analytic continuation into the complex plane could subsequently be performed.

However, a more promising strategy is to extend the variational principle to the $(2+n)$-point function itself. This requires the consistent amputation of the dressed nucleon propagators and automatically leads to agreement with first order perturbation theory for small coupling constants. In a future publication we will show that such an extension is indeed possible at least for the simple model field theory which we have considered up to now. Further corrections in powers of the difference between the exact effective action and the trial action can then be calculated in a similar way as for the polaron problem [17]. The variational principle in the particle representation of field theory thus leads to a systematic sequence of nonperturbative approximations.

# Appendix : Weight function for $\xi(\tau)$

Here we give explicit expressions for the weight function $\rho(E)$ which determines the function $\xi(\tau)$ via

$$\xi(\tau) = \int_{E_0}^{\infty} dE\, \rho(E)\, e^{-E\tau} \tag{A.1}$$

for the various parametrizations of the retardation function and derive some general properties of $\rho$ and $\xi$ in terms of the associated retardation function.

We start with Feynman's profile function. Although it does not have the analytic structure in the complex $E$-plane which we we have assumed it is nevertheless possible to apply Eq. (64). This gives

$$\rho_F(E) = \frac{v^2 - w^2}{4v^3}\, \delta(E - v)\,, \tag{A.2}$$

and obviously yields Eq. (50) when substituted into Eq. (A.1).

For profile functions $A(E)$ which have a cut running along the imaginary $E$-axis from $iE_0$ to infinity we introduce

$$G(E) = \mp \operatorname{Im} A(iE \pm \epsilon) \tag{A.3}$$
$$R(E) = \operatorname{Re} A(iE \pm \epsilon)\,. \tag{A.4}$$

Then Eq. (64) for the weight function reads

$$\rho(E) = \frac{1}{2\pi E^2}\, \frac{G(E)}{R^2(E) + G^2(E)} \tag{A.5}$$

and we have to study the functions $G(E)$ and $R(E)$.

For the 'improved' parametrization we write the profile function (15) as

$$A_I(E) = 1 - \frac{v^2 - w^2}{w}\, \frac{1}{E^2}\left[(w - iE)\ln\left(1 - \frac{iE}{w}\right) + (w + iE)\ln\left(1 + \frac{iE}{w}\right)\right] \tag{A.6}$$

and for $E > 0$ we can read off the following expressions

$$G_I(E) = \pi \frac{v^2 - w^2}{w}\, \frac{E - w}{E^2}\, \Theta(E - w)\,, \tag{A.7}$$

$$R_I(E) = 1 + \frac{v^2 - w^2}{w}\, \frac{1}{E^2}\left[(w + E)\ln\left(1 + \frac{E}{w}\right) - (E - w)\ln\left|\frac{E}{w} - 1\right|\right]\,. \tag{A.8}$$

Note that $G_I(E) \geq 0$ and that near threshold

$$G_I(E) \xrightarrow{E \to w} \pi \frac{v^2 - w^2}{w^3}\, (E - w) \tag{A.9}$$

$$R_I(E) \xrightarrow{E \to w} 1 + 2\frac{v^2 - w^2}{w^2}\, \ln 2\,. \tag{A.10}$$

For the 'extended' parametrization we write Eq. (88) as

$$A_E(E) = 1 - \frac{2C_1}{E^2}\left[(w_1 - iE)\ln\left(1 - \frac{iE}{w_1}\right) + (w_1 + iE)\ln\left(1 + \frac{iE}{w_1}\right)\right.$$
$$\left. + 2\, C_2\sqrt{\pi}\left(\sqrt{w_2 + iE} + \sqrt{w_2 - iE} - 2\sqrt{w_2}\right)\right]\,, \tag{A.11}$$

and obtain

$$G_E(E) = \frac{2\pi C_1}{E^2}\left[(E-w_1)\,\Theta(E-w_1) - \frac{2C_2}{\sqrt{\pi}}\sqrt{E-w_2}\,\Theta(E-w_2)\right], \tag{A.12}$$

$$R_E(E) = 1 + \frac{2C_1}{E^2}\left[(w_1+E)\ln\left(1+\frac{E}{w_1}\right) - (E-w_1)\ln\left|\frac{E}{w_1}-1\right|\right.$$
$$\left.+ 2C_2\sqrt{\pi}\left(\sqrt{w_2+E}+\sqrt{w_2-E}\,\Theta(w_2-E) - 2\sqrt{w_2}\right)\right]. \tag{A.13}$$

In particular,

$$G_E(E) \xrightarrow{E\to w_2} -4\sqrt{\pi}\frac{C_1 C_2}{w_2^2}\sqrt{E-w_2} \tag{A.14}$$

$$R_E(E) \xrightarrow{E\to w_2} 1 + \frac{2C_1}{w_2^2}\left[(w_1+w_2)\ln\left(1+\frac{w_2}{w_1}\right) + (w_1-w_2)\ln\left(1-\frac{w_2}{w_1}\right)\right.$$
$$\left.-2(2-\sqrt{2})C_2\sqrt{\pi w_2}\right]. \tag{A.15}$$

Note that $G_E(E)$ is negative at least between the first and the second branch point.

Some general properties of the functions $G(E), R(E), A(E)$ and $f(\sigma)$ are worthwhile to be noticed and illustrated by the particular examples we have given above. The real part $R(E)$ is, of course, related to $G(E)$ by a dispersion relation

$$R(E) = 1 + \frac{1}{\pi}\,\mathrm{P}\int_{-\infty}^{+\infty} dE' \frac{G(E')}{E'-E} \tag{A.16}$$

which can be further simplified by noting that $G(E')$ is odd and therefore $R(E)$ is even. The profile function itself may be expressed by $G(E)$ as

$$A(E) = 1 + \frac{2}{\pi}\int_{E_0}^{\infty} dE' \frac{E'}{E'^2+E^2}\,G(E'). \tag{A.17}$$

At small $E$ this becomes

$$A(E) \xrightarrow{E\to 0} 1 + \frac{2}{\pi}\int_{E_0}^{\infty} dE'\,\frac{1}{E'}\,G(E') - E^2\,\frac{2}{\pi}\int_{E_0}^{\infty} dE'\,\frac{1}{E'^3}\,G(E') + ... \tag{A.18}$$

If $G(E')$ is negative near threshold (as is the case for the 'extended' parametrization) then the coefficient multiplying the $-E^2$ term may become negative as well. The profile function then will *rise* at small $E$ from the value

$$A(0) = 1 + \frac{2}{\pi}\int_{E_0}^{\infty} dE\,\frac{1}{E}\,G(E). \tag{A.19}$$

We can also express the inverse of the profile function by the discontinuity of $1/A(iE)$ across the cut, i.e. by $\rho(E)$. In particular, at $E=0$ we find

$$A(0) = \left(1 - 4\int_{E_0}^{\infty} dE\, E\, \rho(E)\right)^{-1}, \tag{A.20}$$

assuming that $A(E)$ has no zeroes in the upper half-plane. This can be checked numerically by comparing Eq. (A.20) with either Eq. (A.19) or with the explicit analytic expression, if available.

For the 'extended' parametrization we have verified the equality to be better then 1 part in $10^8$ for all coupling constants.

It is also easy to see that the retardation function $f(\sigma)$ is the Laplace transform of $E^2 G(E)$. Inversion then gives a direct relation between $G(E)$ and the retardation function

$$G(E) = \frac{1}{E^2} \int_{-\infty}^{+\infty} d\sigma\, f(i\sigma + \epsilon)\, e^{iE\sigma}, \qquad E > 0. \tag{A.21}$$

The large $E$-behaviour of $G(E)$ is linked to the the small $\sigma$-behaviour of $f(\sigma)$ and vice versa. For example, the $1/\sigma^2$-singularity of the realistic retardation functions leads to the asymptotic $1/E$-behaviour of $G(E)$ which is observed in Eqs. (A.7) and (A.12). Conversely, Eq. (A.21) may be used to show that a retardation function which asymptotically behaves like

$$f(\sigma) \stackrel{\sigma \to \infty}{\Longrightarrow} \frac{C}{\sigma^\gamma} e^{-E_0 \sigma} \tag{A.22}$$

leads to a behaviour near threshold like

$$G(E) \stackrel{E \to E_0}{\Longrightarrow} \frac{2\pi C}{E_0^2\, \Gamma(\gamma)} (E - E_0)^{\gamma - 1}\, \Theta(E - E_0). \tag{A.23}$$

With Eq. (A.5) this implies that the corresponding weight function has the threshold behaviour

$$\rho(E) \stackrel{E \to E_0}{\Longrightarrow} \frac{C}{E_0^4\, \Gamma(\gamma)} \frac{1}{R^2(E_0)} (E - E_0)^{\gamma - 1}\, \Theta(E - E_0), \tag{A.24}$$

where $R(E_0)$ is the real part of the profile amplitude at the first branch point.

Finally we give the asymptotic expansions for $\xi(\tau)$ for $\tau$ being either small or large. Assuming the canonical $\sigma^{-2}$-behaviour of the retardation function at small $\sigma$ one finds in the first case

$$\xi(\tau) \stackrel{\tau \to 0}{\Longrightarrow} \xi(0) - \left(1 - \frac{1}{A_0}\right) \frac{\tau}{4} - \xi_2\, \tau^2\, \ln \tau + \mathcal{O}(\tau^2) \tag{A.25}$$

where

$$\xi_2 = \frac{1}{2} \lim_{\sigma \to 0} \sigma^2 f(\sigma). \tag{A.26}$$

The logarithmic term is due to the $1/E^3$-fall off of $\rho(E)$ for large $E$ which does not allow a naive expansion of the exponential in Eq. (A.1). It produces a cut for $\xi(\tau)$ on the negative real $\tau$-axis. The linear term in $\tau$ has been expressed by $A_0$ with the help of Eq. (A.20). This is in agreement with the relation (48) between $\xi(\tau)$ and the pseudotime and the small-$\tau$-behaviour (19) of the latter.

For $\tau \to \infty$ the threshold behaviour (A.24) of the weight function is relevant and leads to

$$\xi(\tau) \stackrel{\tau \to \infty}{\Longrightarrow} \frac{C}{E_0^4\, R^2(E_0)} \frac{e^{-E_0 \tau}}{\tau^\gamma}. \tag{A.27}$$

For purely imaginary $\tau$ (which is needed in the analytic continuation) this results in a relatively slow and oscillating decrease at infinity.